\documentclass[crcready,onecolumn]{iosart2x}
\makeatletter
\let\numberlines@hook\relax
\makeatother

\usepackage{dcolumn}

\newcolumntype{d}[1]{D{.}{.}{#1}}

\firstpage{1}
\lastpage{5}
\volume{1}
\pubyear{0000}

\begin{document}
\begin{frontmatter} 

%
\title{Impact of crystallite size on the performance of a beryllium reflector}

\runningtitle{Proceedings of ICANS-XXIII}

\author[A,B]{\fnms{Douglas D.} \snm{DiJulio}\ead[label=e1]{douglas.dijulio@esss.se}%
\thanks{Corresponding author. \printead{e1}.}}
\author[A]{\fnms{Yong Joong} \snm{Lee}}
and
\author[A]{\fnms{Gunter} \snm{Muhrer}}
\address[A]{European Spallation Source ESS ERIC, SE-221 00 Lund, Sweden}
\address[B]{Department of Physics, Lund University, SE-221 00 Lund, Sweden}

\begin{abstract}
Beryllium reflectors are used at spallation neutron sources in order to enhance the low-
energy flux of neutrons emanating from the surface of a cold and thermal moderator. The
design of such a moderator/reflector system is typically carried out using detailed Monte-
Carlo simulations, where the beryllium reflector is assumed to behave as a poly-
crystalline material. In reality, however, inhomogeneities in the beryllium
could lead to discrepancies between the performance of the actual system when compared
to the modeled system. The dependence of the total cross section in particular on crystallite size, in
the Bragg scattering region, could influence the reflector performance, and if such effect
is significant, it should be taken into account in the design of the moderator/reflector
system. In this paper, we report on the preliminary results of using cross-section libraries,
which include corrections for the crystallite size effect, in spallation source neutronics calculations.
\end{abstract}

\end{frontmatter}

\section{Introduction}
The usage of beryllium as an effective neutron reflector at a spallation neutron source has been known since
the early days of spallation neutron source development \cite{Carpenter77,Carpenter81,Kiyanagi87}. The European Spallation
Source \cite{Garoby18} (ESS), currently under construction in Lund, Sweden, will also employ such a reflector material 
to aide in the enhancement of the neutron production from the novel high-brightness cold and thermal neutron moderators
\cite{Zanini19}. To predict the behavior of such a system, it is common practice to carry out detailed Monte-Carlo simulation studies, using
for example such codes such as MCNP6 \cite{MCNP} or PHITS \cite{PHITS}, where the beryllium is treated as a perfect polycrystalline
material. However, inhomogeneities and impurities in the beryllium could lead to discrepancies between the performance of the actual
system when compared to the modeled system. Such a discrepancy was observed for example in a previous study
which compared the simulated performance of a nitrogen-cooled beryllium reflector-filter to the Monte-Carlo simulated performance \cite{Muhrer07}.
It was found in that study, that by manually adjusting the beryllium thermal neutron cross-section library used by the simulations to account for
crystallite effects, good agreement between simulation and experiment could be achieved. \\
\indent The effect of the crystalline state, in particular the size of the crystallites, on the neutron transport through a material was reported in early
neutron transmission measurements \cite{Whitaker39,Fermi47}. This particular effect, know as extinction \cite{Sabine85}, refers to
the re-scattering of the Bragg reflected beam back into the direction of the transmitted beam. This results in an increased observed transmission
of the primary beam and the effect is proportional to the size of the crystallites.  An interesting recent development is that it has
been possible to combine the theoretical treatment of extinction presented in \cite{Sabine85} with Bragg-edge transmission imaging \cite{Sato11,Sato17}
in order to deduce information regarding the crystallite sizes in a material. \\
\indent The current study builds upon the work first presented in \cite{Muhrer07} where the crystallite effects were proposed as one
of the origins for the discrepancy between the simulations and measurements. We aim to quantify the effect of extinction due to crystallites
in beryllium in this work and will not cover the effects of preferred orientation here. In the following, we first give an introduction of how
Bragg neutron scattering is included in thermal neutron scattering simulation libraries, then present how we introduce the extinction effects
into this process, and finally show some preliminary results on the impact of the calculated brightness of the moderators at ESS. 
\section{Methodology}
\begin{figure}[t]
\includegraphics[width=140mm]{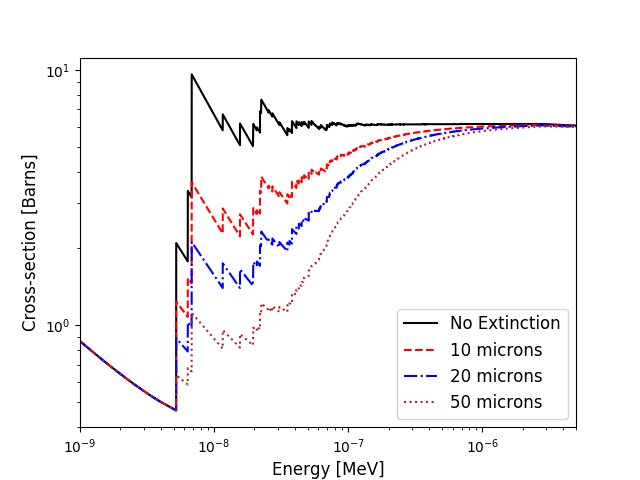}
\caption{Neutron cross-section for beryllium as a function of different crystallite sizes.}
\end{figure}
\begin{figure}[t]
\includegraphics[width=70mm]{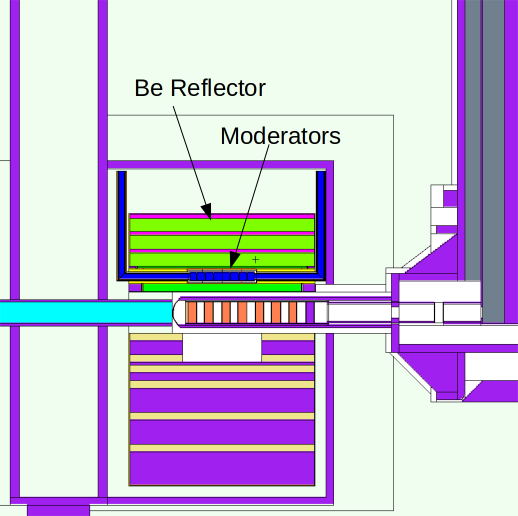}
\caption{MCNP6 ESS model used for the calculation of the cold neutron brightness.}
\end{figure}
For a material such as beryllium, the Bragg scattering from the planes in the crystal results in a jagged neutron cross-section as a function of energy.
In thermal neutron scattering cross-section calculation codes, such as NJOY \cite{MacFarlane10} and NCrystal \cite{Cai19} this effect is included
by assuming a powder average, i.e. small crystallites of random orientation, and the coherent elastic scattering cross-section is given by the following,
\begin{equation}
\sigma_{el}^{coh}(\lambda)=\frac{\lambda^{2}}{2V}\sum_{hkl}d_{hkl}|F_{hkk}|^{2},
\end{equation}
where $V$ is the crsytal structure unit cell volume, $d_{hkl}$ is the lattice plane spacing, $hkl$ are the Miller indices, $\lambda$ is the wavelength, and $F_{hkl}$ is the structure
factor. The extension of this formula to include extinction effects is rather straightforward, as shown in \cite{Sato11}, and results in,
\begin{equation}
\sigma_{el}^{coh}(\lambda)=\frac{\lambda^{2}}{2V}\sum_{hkl}d_{hkl}|F_{hkk}|^{2}E_{hkl}(\lambda,F_{hkl}),
\end{equation}
and $E_{hkl}(\lambda,F_{hkl})$ is the extinction factor, given by,
\begin{equation}
E_{hkl}(\lambda,F_{hkl}) = E_Bsin^2\theta_{hkl}+E_Lcos^2\theta_{hkl},
\end{equation}
with
\begin{equation}
  \theta_{hkl}=arcsin\left(\frac{\lambda}{2d_{hkl}}\right), 
\end{equation}
and where $E_B$ and $E_L$ represent the Bragg and Laue components of the scattering and are given by,
\begin{equation}
  E_B=\frac{1}{\sqrt{1+x}},
\end{equation}
\begin{equation}
  E_L=1-\frac{x}{2}+\frac{x^2}{4}-\frac{5x^3}{48}+... \quad \textrm{for} \quad x\le1,
\end{equation}
\begin{equation}
  E_L=\sqrt{\frac{2}{x\pi}}\left[1-\frac{1}{8x}-\frac{3}{128x^2}-\frac{15}{1024x^3}-...\right] \quad \textrm{for} \quad x>1, 
\end{equation}
and
\begin{equation}
x=S^2\left(\frac{\lambda F_{hkl}}{V} \right)^2, 
\end{equation}
where $S$ is the crystallite size with units of length. The overall effect is that as $S$ increases, the coherent elastic scattering
cross-section $\sigma_{el}^{coh}(\lambda)$ decreases, thus increasing the mean free path of a neutron in the material. \\
\indent Nuclear data libraries were then generated for different crystallite sizes using a modified NJOY code system and based
on the beryllium input file data given in \cite{MacFarlane94}. The calculation of the coherent elastic scattering component
was carried out using a NCrystal-NJOY framework developed by Jose Ignacio Marquez Damian and
available online at \cite{Damian}. The calculation was modified to include the extinction effect described above.
\section{Results and Discussion}
The cross-sections as a function of energy for various crystallite sizes are shown in Fig. 1. As seen in the figure, the increase in crystallite size
reduces the cross-section in the Bragg scattering energy regime. The observed effect agrees with the trend reported in earlier studies on the crystallite effect
in beryllium \cite{Hausner}.\\
\indent After creation of the neutron scattering libraries, we used them to calculate the effect of the crystallite size in the beryllium
reflector on the cold neutron brightness. Fig. 2 shows the model that was used for these calculations, which is an early design of the butterfly 1 concept \cite{Zanini19}.
The procedure for calculating the brightness followed the description as given in Ref. \cite{Zanini19}. \\
\indent Table 1 shows the average relative effects on the cold neutron brightness when using the extinction corrections for the 42 beam ports at ESS. The data are presented
relative to the case when using the neutron scattering library with no extinction corrections, as shown in Fig 1. The overall trend is that an increasing crystallite size
results in a lower average brightness.\\
\indent In addition to having the possibility to more accurately predict the neutronic performance of the beryllium with different crystallite sizes, the ability to link these effects
to the performance of the reflector could be used as valuable input in the beryllium selection process. For example, it could be possible to relax restrictions on the crystallite size if a small
impact on the neutronic performance is deemed acceptable, which could translate into a cost savings for the facility. An important point to also mention is that the crystallite sizes
indicated in this work may not reflect the grain sizes in the material. The possible presence of a sub-grain structure would mean that the crystallites would have smaller dimensions than the grain sizes.
For the above mentioned reasons, a detailed experimental characterization of the beryllium would provide a valuable benchmark to the work presented here. 
\begin{table}[t]
  \caption{The average effects on the ESS cold neutron brightness for the 42 beam ports when using the thermal neutron scattering libraries.
  The presented data are relative to the results when using the library with no extinction effects, as shown in Fig. 1}
\begin{center}
\begin{tabular}{|c|c|}
\hline
Crystallite Size  & Average Relative Effect     \\
\hline
No Extinction   & 100\%  \\
10 microns   & 97.5\%  \\
20 microns  & 95.1\%    \\
50 microns  & 92.2\%   \\
\hline
\end{tabular}
\end{center}
\label{default}
\end{table}%

\section{Conclusion}
In summary, we have demonstrated how to include extinction effects directly into the thermal neutron data libraries used
for Monte-Carlo simulations. We have used this method to study the effect of crystallite sizes on the performance
of a beryllium reflector at a spallation neutron source. Future work will entail comparisons with detailed neutron
transmission measurements and looking into the possibilities for including prefered orientation effects into the
calculation process.

\section{Acknowledgments}
The authors would like to thank Luca Zanini for valuable discussions on this topic and for providing the MCNP6 model for the brightness calculations.
The authors would also like to thank Thomas Kittelmann for fruitful discussions and NCrystal support related to aspects of this work.


\begin{thebibliography}{0}

\bibitem{Carpenter77} J. M. Carpenter, Nucl. Instrum. and Meth. \textbf{145} (1977) 91

\bibitem{Carpenter81} J. M. Carpenter et al., Nucl. Instrum. and Meth. \textbf{189} (1981) 485

\bibitem{Kiyanagi87} Y. Kiyanagi J. Nucl. Sci. Tech. \textbf{24} (1987) 490

\bibitem{Garoby18} R. Garoby et al., Phys. Scr. \textbf{93} (2018) 014001

\bibitem{Zanini19} L. Zanini et al., Nucl. instrum. and Meth. Phys. Res. A \textbf{925} (2019) 33

\bibitem{MCNP} C. J. Werner, et al., "MCNP6.2 Release Notes", Los Alamos National Laboratory, report LA-UR-18-20808 (2018).

\bibitem{PHITS} T. Sato et al., J. Nucl. Sci. Technol. \textbf{55}, (2018) 684

\bibitem{Muhrer07} G. Muhrer et al., Nucl. Instrum. and Meth. Phys. Res. A, \textbf{578} (2007) 463

\bibitem{Whitaker39} M. D. Whitaker and H. G. Bayer, Phys. Rev. \textbf{55} (1939) 1101

\bibitem{Fermi47} E. Fermi, W. J. Sturm and R. G. Sachs, Phys. Rev. \textbf{71} (1947) 589

\bibitem{Sabine85} T. M. Sabine, Aust. J. Phys, \textbf{38} 1985 507

\bibitem{Sato11} H. Sato, T. Kamiyama, and Y. Kiyanagi, Materials Transactions \textbf{52} (2011) 1294

\bibitem{Sato17} H. Sato et al., Phys. Procedia \textbf{88} (2017) 322

\bibitem{MacFarlane10} R. E. MacFarlane and A. C. Kahler, Nuclear Data Sheets \textbf{111} (2010) 2739

\bibitem{Cai19} X.-X. Cai and T. Kittelmann Comp. Phys. Comm. \textbf{246} (2020) 106851

\bibitem{MacFarlane94} R. E. MacFarlane, New thermal neutron scattering files for ENDF/B-VI release 2, Technical Report LA-12639-MS

\bibitem{Damian} https://gist.github.com/marquezj

\bibitem{Hausner} H. H. Hausner, Beryllium: It's Metallurgy and Properties, University of California Press, Berkeley and Los Angeles, 1965
\end{thebibliography}
\end{document}